# All-dielectric nanolaser

D. G. Baranov,[1,2] A. P. Vinogradov,[1,2,3] and A. A. Lisyansky[4,5,a)]

[1]*Moscow Institute of Physics and Technology, 9 Institutskiy per., Dolgoprudny 141700, Russia*

[2]*All-Russia Research Institute of Automatics, 22 Sushchevskaya, Moscow 127055, Russia*

[3]*Institute for Theoretical and Applied Electromagnetics, 13 Izhorskaya, Moscow 125412, Russia*

[4]*Department of Physics, Queens College of the City University of New York, Queens, New York 11367, USA*

[5]*The Graduate Center of the City University of New York, New York, New York 10016, USA*

We demonstrate theoretically that a subwavelength spherical dielectric nanoparticle coated with a gain shell forms a nanolaser. Lasing modes of such a nanolaser are associated with the Mie resonances of the nanoparticle. We establish a general condition for the lasing threshold and show that the use of high refractive index dielectric media allows for lasing with a threshold significantly lower than that for nanolasers incorporating lossy plasmonic metals.

A nanolaser is a novel type of quantum device that enables the generation of a coherent electromagnetic field at the subwavelength scale.[1-4] In order to localize the electromagnetic field of a laser mode at the nanoscale, one employs surface plasmons instead of photons because the wavelength of a plasmon is shorter than that of a photon at the same frequency. The nanolaser can be several wavelengths in length[5] producing a beam of light like a usual laser or be of subwavelength size radiating like a multipole. In the latter case, it is referred to as a spaser.[1] Basically, a spaser consists of a plasmonic nanoparticle (NP) coupled to a gain medium pumped by an external energy source. A serious problem that hinders the experimental realization of spasers is that the extreme loss in plasmonic metals raises the lasing threshold of spasers.[6, 7] In addition, the effects of nonlocality and spatial dispersion generally increase the lasing threshold.[8] Attempts have been made to lower the high gain required for generation in plasmonic-based nanolasers using a special design for core-shell cavities.[9, 10] However, the resulting threshold is still high. In particular, use of the optimal core-shell design[9] allows one to lower threshold gain to values as low as $4 \cdot 10^4$ cm$^{-1}$, while a characteristic gain provided by organic dyes is of order of several $10^3$ cm$^{-1}$.

Semiconductor media exhibiting high refractive indices in the optical range, such as Si, Ge, and GaAs, allow for a realization of subwavelength spherical resonators without the use of lossy metals. Note that in the near infrared region these materials should be considered as dielectrics rather than semiconductors, which they are in the optical region. It was predicted theoretically that spherical silicon NPs exhibit a strong low-loss magnetic response in the optical

---

[a)] Corresponding author: lisyansky@qc.edu

frequency range.[11] Thanks to the progress in the fabrication of silicon nano-sized spherical particles,[12, 13] the predicted strong magnetic response of dielectric particles associated with Mie resonances has become experimentally accessible. Very recently, it has been proposed that the use of a gain medium can further enhance the magnetic response of silicon particles.[14] Such a high-quality resonator may serve as a platform for the design of a quantum generator of a coherent near-field. As with a spaser, in which a plasmon mode localized on a metallic NP couples to the gain medium, in the case of a high-permittivity spherical dielectric particle surrounded by an active material, the near-field of a certain Mie resonance should couple to the gain medium.

Scattering of light from spherical particles with gain, or "negative absorption", has been studied long ago in a number of works.[15, 16] Although peculiarities in the scattering spectrum by a particle were found, their connection to the lasing modes of an amplifying particle has not been discussed. In those works, the response of the amplifying nanoparticle has been studied for a discreet set of gain values, while a scattering singularity, which gives rise to the laser generation, occurs at one point in the gain parameter space. Later, the connection between these singularities and lasing has been established.[17] After that, lasing in dielectric cavities through Mie resonances has been extensively studied.[18-20] Mie resonances giving rise to lasing considered in those studies are the well-known whispering gallery mode (WGM) resonances. Resonances supported by subwavelength semiconductor NPs should be distinguished from the WGM resonances, which occur in particles with sizes larger than the free space wavelength.[21] Unlike WGM lasers, the semiconductor NP based nanolaser theoretically described in our work operates at *low order Mie resonances* which occur at free space wavelength larger than the NP dimension. This is the main feature of our work.

In this paper, we investigate self-oscillations of a core-shell dielectric nanostructure containing an active medium with inverse population. We show that a lasing transition can be observed beyond a certain (threshold) level of gain appearing as a singularity of the scattering cross section. We find the set of lasing modes of the dielectric nanolaser and, using actual material parameters for high refractive-index media, compare the threshold characteristics of dielectric-based nanolasers with those of metallic-based spasers.

Although the laser generation is essentially a nonlinear process, the threshold pumping for lasing can be found within a linear approximation.[22, 23] The transition to lasing emerges as a non-trivial solution of Maxwell's equations in the absence of an incident field. The amplitude of this *self-oscillation* is determined by an excess of the pump rate over the pumping threshold. At the threshold, the amplitude of the laser self-oscillation is equal to zero. Therefore, for pump powers below and at the threshold, the problem can be treated linearly. Assuming the $e^{i\omega t}$-time dependence of the electromagnetic fields we describe the gain medium by the Lorentzian permittivity with the negative imaginary part due to population inversion created by an external pump:[24, 25]

$$\varepsilon_{gain}(\omega) = \varepsilon_0 + D_0 \frac{2\omega_0 \gamma}{\omega^2 - \omega_0^2 + 2i\omega\gamma}, \tag{1}$$

where $\varepsilon_0$ is the background permittivity of the gain medium in the absence of pumping, $\omega_0$ is the emission frequency, $\gamma$ is the emission linewidth, and $D_0$, which equals $|\mathrm{Im}\,\varepsilon_{gain}|$ at the emission frequency, characterizes the gain strength.

The geometry of the problem is illustrated in the inset of Fig. 1. The system under study is a core-shell spherical NP consisting of an inner high-permittivity core coated with the gain shell. The total radius of the NP is $R$, which is less than the free-space wavelength, the radius of the core is $r$. Figure 1 shows the frequency dependence of the scattering cross section of the core-shell particle with no loss or gain, i.e., at $D_0 = 0$. In the optical frequency range, the core-shell particle exhibits a series of resonances that are represented mainly by dipole and quadrupole terms.

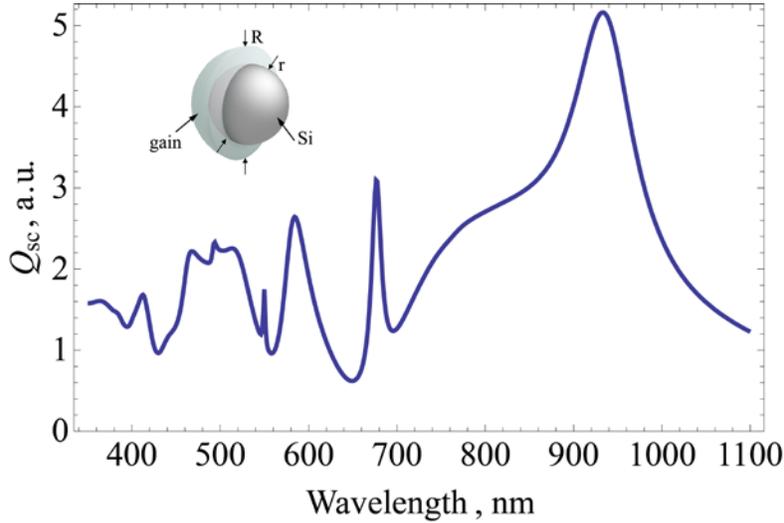

FIG. 1. The scattering cross section (normalized by the geometrical cross section) of a passive ($D_0 = 0$) core-shell NP with a silicon core of radius $r$ = 120 nm. The radius of the whole NP is $R$ = 1.7$r$ and the background shell permittivity is $\varepsilon_0 = 2$. Here and below, the permittivity of silicon is taken from Ref. 26. Inset: schematic of the core-shell dielectric nanolaser.

In order to find the electromagnetic self-oscillation of a core-shell spherical particle, we look for a non-zero solution of Maxwell's equations in absence of an incident field. Following the standard approach, we decompose electric and magnetic fields in the dielectric core in a series of spherical vector functions $\mathbf{M}_{o1n}^{(1)}$, $\mathbf{M}_{e1n}^{(1)}$, $\mathbf{N}_{o1n}^{(1)}$ and $\mathbf{N}_{e1n}^{(1)}$, expressions for which may be found in Ref. 27:

$$\mathbf{E}_{core} = \sum_{n=1}^{\infty} E_n \left( c_n \mathbf{M}_{o1n}^{(1)} - i d_n \mathbf{N}_{e1n}^{(1)} \right),$$

$$\mathbf{H}_{core} = -\frac{k\sqrt{\varepsilon_{core}}}{\omega} \sum_{n=1}^{\infty} E_n \left( d_n \mathbf{M}_{e1n}^{(1)} + i c_n \mathbf{N}_{o1n}^{(1)} \right), \quad (2)$$

where $E_n = i^n (2n+1)/[n(n+1)]$ and $k = \omega/c$. For electric and magnetic fields in the gain shell we have

$$\mathbf{E}_{shell} = \sum_{m=1}^{\infty} E_n \left( f_n \mathbf{M}_{o1n}^{(1)} - i g_n \mathbf{N}_{e1n}^{(1)} + v_n \mathbf{M}_{o1n}^{(2)} - i w_n \mathbf{N}_{e1n}^{(2)} \right),$$

$$\mathbf{H}_{shell} = -\frac{k\sqrt{\varepsilon_{shell}}}{\omega} \sum_{m=1}^{\infty} E_n \left( g_n \mathbf{M}_{e1n}^{(1)} + i f_n \mathbf{N}_{o1n}^{(1)} + w_n \mathbf{M}_{e1n}^{(2)} + i v_n \mathbf{N}_{o1n}^{(2)} \right). \quad (3)$$

The fields in the outer region can be represented as

$$\mathbf{E}_{out} = \sum_{m=1}^{\infty} E_n \left( i a_n \mathbf{N}_{e1n}^{(3)} - b_n \mathbf{M}_{o1n}^{(3)} \right),$$

$$\mathbf{H}_{out} = \frac{k}{\omega} \sum_{m=1}^{\infty} E_n \left( i b_n \mathbf{N}_{o1n}^{(3)} + a_n \mathbf{M}_{e1n}^{(3)} \right). \quad (4)$$

The field in the external region contains only terms corresponding to the outgoing solution, which implies an absence of incident radiation. This field is non-zero if at least one of $a_n$ or $b_n$ coefficients is non-zero for some $n$. By imposing boundary conditions at the interfaces between high-permittivity and gain media, and between the gain medium and the vacuum, we arrive at a system of linear equations, allowing us to determine unknown coefficients in series (2), (3), and (4):

$$\begin{aligned}
&f_n m_1 \psi_n (m_2 x) - v_n m_1 \chi_n (m_2 x) - c_n m_2 \psi_n (m_1 x) = 0, \\
&w_n m_1 \chi'_n (m_2 x) - g_n m_1 \psi'_n (m_2 x) + d_n m_2 \psi'_n (m_1 x) = 0, \\
&v_n \chi'_n (m_2 x) - f_n \psi'_n (m_2 x) + c_n \psi'_n (m_1 x) = 0, \\
&g_n \psi_n (m_2 x) - w_n \chi_n (m_2 x) - d_n \psi_n (m_1 x) = 0, \\
&-a_n m_2 \xi'_n (y) - g_n \psi'_n (m_2 y) + w_n \chi'_n (m_2 y) = 0, \\
&m_2 b_n \xi_n (y) - m_2 \psi_n (y) + f_n \psi_n (m_2 y) - v_n \chi_n (m_2 y) = 0, \\
&-a_n \xi_n (y) - g_n \psi_n (m_2 y) + w_n \chi_n (m_2 y) = 0, \\
&b_n \xi'_n (y) + f_n \psi'_n (m_2 y) - v_n \chi'_n (m_2 y) = 0,
\end{aligned} \quad (5)$$

where $x = kr$, $y = kR$, $m_1 = \sqrt{\varepsilon_{core}}$ and $m_2 = \sqrt{\varepsilon_{shell}}$, $\psi(x)$, $\chi(x)$, and $\xi(x)$ are the Riccati-Bessel functions defined as $\psi(x) = \sqrt{\pi x/2} J_{n+1/2}(x)$, $\xi(x) = \sqrt{\pi x/2}\left[J_{n+1/2}(x) + iY_{n+1/2}(x)\right]$ and $\chi(x) = -\sqrt{\pi x/2} Y_{n+1/2}(x)$.[27] Note that system (5) is composed of two independent linear systems for the two sets of coefficients $\{a_n, g_n, w_n, d_n\}$ and $\{b_n, f_n, v_n, c_n\}$, respectively. The system for coefficients $\{a_n, g_n, w_n, d_n\}$ determines the electric modes of the spherical resonator, while the other system determines the magnetic modes. To be specific, we will study the "magnetic" subsystem of the linear equations and the corresponding magnetic laser modes. The treatment of electric modes is similar.

The condition for the existence of a non-zero solution to a uniform system of linear equations is the equality of its determinant to zero:

$$m_2 \xi(y)\left[\psi'_n(m_2 y) - \beta_n \chi'_n(m_2 y)\right] - \xi'_n(y)\left[\psi_n(m_2 y) - \beta_n \chi_n(m_2 y)\right] = 0, \tag{6}$$

where

$$\beta_n = \frac{m_2 \psi_n(m_1 x) \psi'_n(m_2 x) - m_1 \psi_n(m_2 x) \psi'_n(m_1 x)}{m_2 \chi'_n(m_2 x) \psi_n(m_1 x) - m_1 \psi'_n(m_1 x) \chi_n(m_2 x)}. \tag{7}$$

Nonlinear complex-valued Eq. (6) determines the lasing threshold, $D_{th}$, as well as the frequency of the self-oscillation. Note that at fixed pumping, below the threshold, Eq. (6) represents the condition for the poles of the Mie coefficients $B_n$.[27] In a problem with an incident field, these coefficients determine the electromagnetic field scattered by the NP. Similarly, the condition for the existence of the electric lasing mode coincides with the condition for the pole of $A_n$ coefficients. Therefore, the lasing modes of a core-shell spherical NP are associated with the poles of the corresponding Mie coefficients. Below the threshold, the scattering singularities have complex-valued frequencies. Specifically, these singularities are located in the frequency lower half-plane; i.e., their frequencies have a negative imaginary part, $\text{Im}\,\omega < 0$, indicating the exponential in time decay of the corresponding quasistationary state.[28] With a gain increase in the system, the poles may reach the real frequency axis on the complex plane, giving rise to a lasing mode.[23, 25]

Let us show that, for a certain value of gain, the pole of the magnetic coefficient, $B_1$, appears at a real-valued frequency. To demonstrate the existence of a self-oscillating solution we tune the emission frequency of the gain medium to the dipole magnetic resonance of the core-shell nanolaser, occurring at approximately $\lambda \approx 940\,\text{nm}$. Figure 2 shows the logarithm of the squared absolute value of the Mie coefficient, $B_1$, as a function of the wavelength and the gain parameter $D_0$. At a certain wavelength, when the gain reaches a value close to $D_0 = 1.0$, $|B_1|^2$ is increased by orders of magnitude (the peak denoted by a black color in Fig. 2). This feature of the graph is a clear indication that a pole of the Mie coefficient has been reached and, therefore, that laser generation should begin. Importantly, the resulting lasing wavelength is not necessarily equal to the gain medium emission wavelength if it was initially tuned to the wavelength of the Mie resonance. This occurs due to an increase of the refractive index of the gain shell induced by the increase in gain that leads to the shift of the "cold" resonance wavelength. With a further increase in gain, $|B_1|^2$ decreases. This behavior, however, does not imply that the lasing has been turned off, because when the threshold is reached, the laser becomes a strongly nonlinear system, and saturation of the gain medium must be taken into account to correctly describe the behavior of the laser above the threshold. In fact, when the gain exceeds the threshold, the pole of the corresponding Mie coefficient moves to the upper half of the complex frequency plane, $\text{Im}\,\omega > 0$.[23] It corresponds to a solution with a time dependence proportional to the diverging exponential factor, $\exp(\text{Im}\,\omega t)$. This exponential growth is suppressed by the saturation of the gain medium when lasing reaches the steady-state regime.

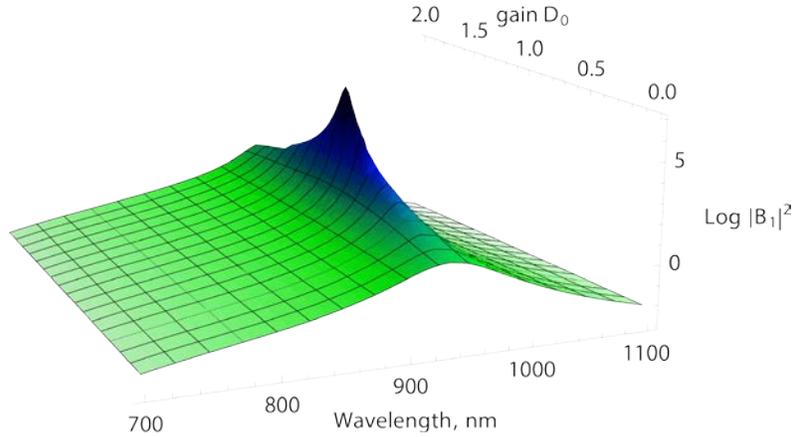

FIG 2. The squared absolute value of the Mie coefficient $B_1$ as a function of the wavelength and gain. A giant increase in $|B_1|$ indicates the pole of the Mie coefficient which is associated with the lasing threshold.

To rigorously show that at a certain level of gain the transition to lasing occurs, we rewrite Eq. (6) in the form of two real-valued equations:

$$\text{Re}\,F(\omega, D_0) = 0, \quad \text{Im}\,F(\omega, D_0) = 0, \qquad (8)$$

where

$$F(\omega, D_0) = m_2 \xi(y)[\psi'_n(m_2 y) - \beta_n \chi'_n(m_2 y)] - \xi'_n(y)[\psi_n(m_2 y) - \beta_n \chi_n(m_2 y)]. \quad (9)$$

In Fig. 3, we depict both real and imaginary parts of $F(\omega, D_0)$ for coefficient $B_1$ calculated at two different values of gain $D_0$ below and above the lasing threshold, respectively. Intersections of the curves $\text{Re}\, F = 0$ and $\text{Im}\, F = 0$ occur on different sides of the real axis for different gains. Since $F(\omega, D_0)$ is continuous, for a certain value of gain $D_{th}$, this intersection occurs exactly on the real axis, indicating the onset of lasing generation.

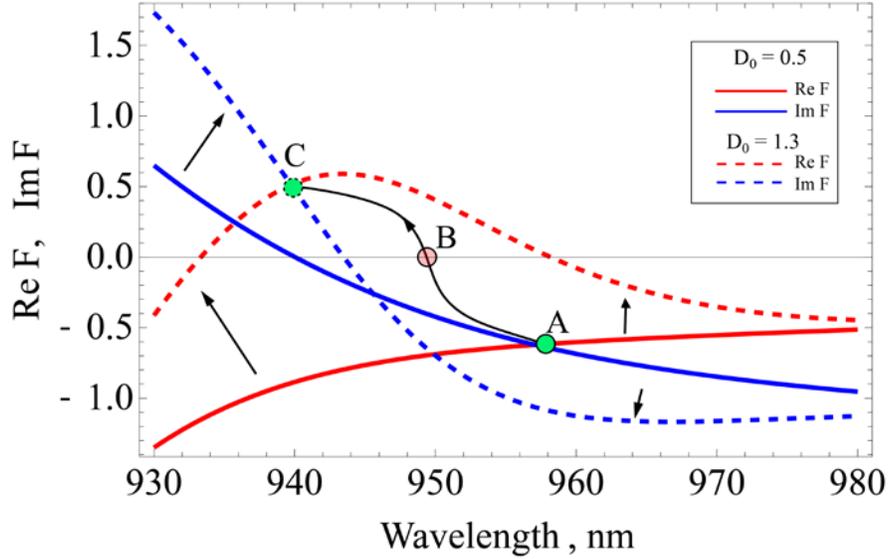

FIG. 3. Real and imaginary parts of $F(\omega, D_0)$ [see Eq. (8)] for values of gain $D_0$ below and above the lasing threshold. The short arrows show the evolution of the two curves with the gain increase. Points A and C show the intersection of $\text{Re}\, F$ and $\text{Im}\, F$ below and above the threshold, respectively. With an increase in gain, point A moves towards the $x$-axis of the plot; it crosses the axis (point B) at a certain value of gain, $0.5 < D_0 < 1.3$. The intersection of $\text{Re}\, F$ and $\text{Im}\, F$ occurring exactly at the point $F = 0$ indicates the pole of the Mie coefficient $B_1$ and the onset of laser generation.

We stress that the lasing modes of the dielectric nanolaser considered in this paper substantially differ from those of WGM lasers. Although WGM lasers have an ultra-low lasing threshold, their sizes are typically of the order of microns or a dozen microns.[21] Indeed, the WGM resonance is a Mie resonance of a dielectric particle that occurs at large $n$. To fulfill the resonance condition for a high-order Mie coefficient, either the wavelength must be small or the particle size must be large.

The gain threshold for lasing for each lasing mode depends strongly on the material parameters and geometry. Table I presents lasing wavelength $\lambda_{las}$, the threshold gain, $D_{th}$, and the corresponding volume gain coefficient, $g_{th} = \text{Im}\,\omega/c\sqrt{\varepsilon_{gain}}$, for the electric and magnetic modes of a core-shell nanolaser corresponding to $n = 1, 2$ with the geometric parameters specified above. In each case, the central frequency $\omega_0$ of the gain medium is adjusted to the resonance frequency of the corresponding lasing mode. At optical frequencies in silicon, the Ohmic loss due to interband transitions is relatively low. Thus, the main loss channel of the dielectric resonator is radiation. Since quadrupoles radiate much less than dipoles, the quadrupole resonances associated with the poles of $A_2$ and $B_2$ have lower lasing thresholds than excitation of dipole electric and magnetic modes.

We now compare the lasing thresholds of the dielectric nanolaser with that of an Au-based spaser. Based on the geometry of spasers,[4] we calculate the response of a NP consisting of the Au core of radius $r = 7\,\text{nm}$ coated by a doped silica shell of thickness $R - r = 15\,\text{nm}$ immersed in water. At optical frequencies, the response of such a NP is dominated by the plasmon resonance at the wavelength $\lambda = 530\,\text{nm}$ (the pole of $A_1$); only the corresponding dipole mode gives rise to lasing. The calculated threshold of the Au-based spaser, $D_{th} = 1.2$, and the corresponding volume threshold gain, $g_{th} = 46 \cdot 10^3\,\text{cm}^{-1}$, are comparable to that of the magnetic dipole mode of the dielectric nanolaser. As can be inferred from Table 1, the thresholds of the quadrupole modes ($A_2$ and $B_2$) of the dielectric dipole nanolaser are significantly lower than those of the spaser. This is especially noticeable for the low-loss magnetic quadrupole mode, $B_2$, whose threshold gain and volume gain coefficient are as low as $D_{th} = 0.35$ and $g_{th} = 11 \cdot 10^3\,\text{cm}^{-1}$. This is also smaller than was previously reported for plasmonic core-shell nanolaser with optimized geometry and material parameters.[9]

Silicon is not the only material that can be used in the design of subwavelength non-metallic nanolasers. In particular, spherical NPs made of other materials with high refractive indexes, such as Ge and GaAs, can be fabricated.[29] From our point of view, however, silicon seems to be the most favorable candidate for the realization of dielectric nanolasers. Due to its electronic bandgap of $\Delta E = 1.1\,\text{eV}$, silicon has a moderate loss with $\varepsilon'' \sim 0.01...0.02$ and a relatively high refractive index of $n \sim 3.5$ in the region in which quadrupole resonances of the core-shell particle appear (wavelengths $600-700\,nm$). GaAs has a larger refractive index at optical frequencies that could allow for lasing modes at smaller particle sizes; however, its absorption is higher by an order of magnitude.[26] Ge also exhibits high absorption at wavelengths up to 1 micron.[26]

TABLE I. Lasing threshold values for different lasing modes of the core-shell dielectric NP and of an Au-based spaser.

| Type of resonance | $A_1$ | $B_1$ | $A_2$ | $B_2$ | $A_1$ (spaser) |
|---|---|---|---|---|---|
| Resonance wavelength, nm | 500 | 945 | 590 | 690 | 530 |
| $D_{th}$ | 1.8 | 1.0 | 0.6 | 0.35 | 1.2 |
| $g_{th}$, cm$^{-1}$ | $76 \cdot 10^3$ | $50 \cdot 10^3$ | $22 \cdot 10^3$ | $11 \cdot 10^3$ | $46 \cdot 10^3$ |

To conclude, we have shown that a subwavelength spherical dielectric NP coupled to a gain medium forms a nanolaser whose lasing modes are associated with the Mie resonances of a NP. We establish the general condition for the lasing threshold, from which one can obtain numerically the level of threshold gain. The thresholds for the electric and magnetic dipole modes of the dielectric nanolaser are comparable to the thresholds for the dipole mode of a spaser with a 7-nm Au core. The thresholds of the quadrupole modes arising in the optical range are significantly lower than the characteristic spaser threshold.

This work was supported by RFBR grants Nos 13-07-92660, 12-02-01093, and 12-02-00407, by Dynasty Foundation, by PSC-CUNY Research Award, and by the NSF under Grant No. DMR-1312707.